\documentclass[prl,a4paper,twocolumn,footinbib,showpacs]{revtex4}

\usepackage[latin1]{inputenc}
\usepackage{amssymb}
\usepackage{amsmath}
\usepackage{amsbsy}
\usepackage{bm}
\usepackage{graphicx}

\begin{document}

\title{Effect of step stiffness and diffusion anisotropy on the\\
meandering of a growing vicinal surface}
\author{Thomas Frisch}
   \altaffiliation{\'Ecole Générale des Ingénieurs de Marseille,
   Technopôle de Château-Gombert,
   Marseille, France}\email{frisch@irphe.univ-mrs.fr}
   \affiliation{%
   Institut de Recherche sur les Phénomènes Hors \'Equilibre,
   UMR 6594, CNRS, Université de Provence, Marseille, France}

   \author{Aberto Verga}\email{Alberto.Verga@irphe.univ-mrs.fr}
   \affiliation{%
   Institut de Recherche sur les Phénomènes Hors \'Equilibre,
   UMR 6594, CNRS, Université de Provence, Marseille, France}

\date{\today}

\begin{abstract}
We study the step meandering instability on a surface characterized
by the alternation of terraces with different properties, as in the
case of Si(001). The interplay between diffusion anisotropy and step
stiffness induces a finite wavelength instability corresponding to a
meandering mode. The instability sets in beyond a threshold value
which depends on the relative magnitudes of the destabilizing flux
and the stabilizing stiffness difference. The meander dynamics is
governed by the conserved Kuramoto-Sivashinsky equation, which
display spatiotemporal coarsening.
\end{abstract}


\maketitle

Molecular beam epitaxy (MBE) is often used to grow nanostructures on
vicinal surfaces of semiconductor and metallic crystals
\cite{stangl04,pimpinelli98,saito98,neel03,nita05}. Under
nonequilibrium growth a very rich variety of crystal surface
morphologies are experimentally observed resulting from the
nonlinear evolution of step bunching and meandering instabilities
\cite{jeong99,politi00,yagi01}. Self-organized patterns arising from
these instabilities may be used for the development of technological
applications such as quantum dots and quantum wells
\cite{shchukin99,brunner02}.

The step meandering instability was originally predicted
theoretically by Bales and Zangwill \cite{bales90} for a vicinal
surface under growth. Its origin comes from the asymmetry between
the descending and ascending currents of adatoms. Nonlinear
extensions of this work have shown that the meander evolution can be
described by amplitude equations showing diverse behaviors:
spatiotemporal chaos governed by the Kuramoto-Sivashinsky equation
in the case of the Erlich-Schwoebel effect  with desorption
\cite{bena93}; nonlinear coarsening in the case of negligible
desorption \cite{pierre-louis98,kallunki00,gillet00}; and
interrupted coarsening when two-dimensional anisotropy is included
\cite{danker03,danker04}. Step meandering on a Si(001) vicinal
surface can also be driven by a drift electromigration current in
the presence of alternating diffusion coefficients, even for
symmetric adatom attachment to the steps \cite{sato03,sato05}.

\begin{figure}
\centering
\includegraphics[width=0.4\textwidth]{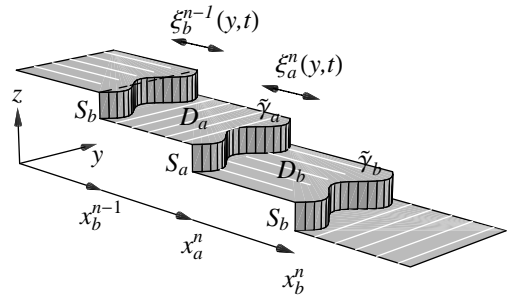}
\caption{Sketch of the Si(001) vicinal surface showing the
alternation of terraces and steps $S_a$ and $S_b$. Lines on terraces
indicate the privileged diffusion directions. $D_a$, $D_b$ and
$\tilde\gamma_a$, $\tilde\gamma_b$ are the surface diffusion and
step stiffness coefficients, respectively; $x^n_{a}$, $x^n_{b}$ and
$\xi^n_{a}(y,t)$, $\xi^n_{b}(y,t)$ are step positions and the
corresponding perturbations.} \label{fig1}
\end{figure}

Recent experiments
\cite{schelling99,schelling00,pascale03,myslivecek02b} on the growth
of Si(001) have revealed the existence of a step bunching
instability and the development of a transverse two-dimensional
complex structure (ripples), possibly reminiscent  of a meandering
instability. We have recently shown that the observed step bunching
instability is due to the interplay between the elastic interactions
and the alternation of the step parameters \cite{frisch05}. In order
to understand the roughening of the Si(001) surface during growth it
would be important to know if step bunching and meandering can arise
simultaneously. So far, no conclusive experimental evidence for the
presence of the Schwoebel effect leading to meandering instability
has been established.

In this Letter, we show that the difference in step stiffness and
diffusion anisotropy induce a meandering instability which can
account for the complex step structures observed in Si(001)
epitaxial growth. We first present the linear two-dimensional
stability analysis of a train of steps using a simplified extension
of the model studied in reference \cite{frisch05}. We show that the
interplay between diffusion anisotropy and the step stiffness effect
under growth conditions induces a finite wavelength instability
which is maximum for the in-phase modes. We present the stability
phase diagram in the parameters space. Our results are complemented
by a weak nonlinear analysis of the step meander which reveals a
coarsening dynamics. Using a simple similarity argument, we show
that the characteristic coarsening exponent is 1/2 and that the
general solution of the CKS equations can be thought as a
superposition of parabolas. Finally we conclude this Letter by
discussing the possible relevance of our model to the experiment on
Si growth and the consequences of the simultaneous existence of
bunching and meandering instabilities.

The Si(001) vicinal surface consists of a periodic sequence of
terraces where rows of $2\times1$ dimerized adatoms (terrace of type
$a$) alternate with $1\times2$ dimerized adatoms  (terrace of type
$b$), as shown in Fig.~\ref{fig1} where we also introduce several
notations. On the reconstructed surface adatoms diffuse
preferentially along dimer rows, giving rise to an anisotropic
diffusion. Therefore, the steps separating the terraces are of two
kinds. The $S_a$ steps are rather straight while the $S_b$ ones are
very corrugated \cite{zandvliet00}. We shall allow each step to have
a different step stiffness coefficient $\tilde\gamma_a$ and
$\tilde\gamma_b$ \cite{bartelt96}. For simplicity we neglect elastic
interactions between steps and assume that the desorption of adatoms
is negligible; we also neglect Erlich-Schwoebel effects. Let us
denote by $x_a^n(y,t)$ and $x_b^n(y,t)$ the positions at time $t$ of
steps $S_a$ and $S_b$ respectively (cf. Fig.~\ref{fig1}). During
growth, the adatom concentrations on each terrace $C_a^n(x,y,t)$ and
$C_b^n(x,y,t)$, obey the following diffusion equations
\cite{burton51}:
\begin{eqnarray}
\label{bcf1}
    D_a \partial^2_{x} C_a^n+D_b\partial^2_{y}C_a^n&=&- F\,,\\
\label{bcf2}
    D_b\partial^2_{x} C_b^n+D_a\partial^2_{y} C_b^n&=&-F \,,
\end{eqnarray}
where $D_a$ and $D_b$ are the diffusion coefficients, and $F$ the
deposition flux. $C_a^n$ and $C_b^n$ are the difference of adatoms
concentrations with respect to the uniform equilibrium concentration
$C_0$ (taken to be the same on both terrace types). We assume for
$C_a^n$ and $C_b^n$, the following boundary conditions:
\begin{eqnarray}
\label{bc1}
    C_a^n(x_b^{n-1})= C_{eq,b}^{n-1} \,  &,&
        \quad
    C_a^n(x_a^n)=C_{eq,a}^n  \, ,\\
\label{bc2}
    C_b^n(x_a^n)=C_{eq,a}^n\, &,&
        \quad
    C_b^n(x_b^n)=C_{eq,b}^n\, ,
\end{eqnarray}
which correspond to instantaneous attachment kinetics, and can be
considered as the simplest ones capturing the main physical effects
(no diffusion along steps and negligible transparency). The adatom
equilibrium concentrations $C_{eq,a}^n$ and $C_{eq,b}^n$ depend on
the step curvatures $\kappa_a^n$ and $\kappa_b^n$ \cite{bena93}:
\begin{equation}
    C_{eq,a}^n = C_0\Gamma_a \kappa_a^n  \, ,
        \quad
    C_{eq,b}^{n} = C_0\Gamma_b \kappa_b^n \, ,
\end{equation}
with $ \Gamma_b= \Omega \tilde\gamma_b/k_BT$ and  $\Gamma_a= \Omega
\tilde\gamma_a/k_BT$ ($\Omega$ is the unit atomic surface, $T$ the
temperature, and $k_B$ the Boltzmann constant). The normal
velocities of each step are $v_a^n=\dot{x}_a^n/(1+(\partial_y
x_a^n)^2)^{1/2}$ and $v_b^n=\dot{x}_b^n/(1+(\partial_y
x_b^n)^2)^{1/2}$, where
\begin{eqnarray}
\label{vab1}
    \dot{x}_a^n &= \Omega&
    \left[
    (D_b\partial_x C_b^n-D_a(\partial_y x_a^n)\partial_y C_b^n)-
    \right.\nonumber\\
    &&\left.
        (D_a\partial_x C_a^n-D_b(\partial_y x_a^n)\partial_y C_a^n)
    \right]_{x=x_a^n}
   \, ,   \\
   \dot{x}_b^n &= \Omega&
    \left[
    (D_a\partial_x C_a^{n+1}-D_b(\partial_y x_b^n)\partial_y C_a^{n+1})-
    \right.\nonumber\\
\label{vab2}
    &&(D_b\left.
    \partial_x C_b^n-D_a(\partial_y x_b^n)\partial_y C_b^n)
    \right]_{x=x_b^n}\,.
\end{eqnarray}
In order to get a nondimensional version of equations
(\ref{bcf1})-(\ref{vab2}), we set the unit of length to be the
initial size of the terrace $l_0$ and the unit of time $l_0^3/(C_0
\Gamma_a \Omega D_a)$. The system is controlled by three independent
positive nondimensional parameters:
\begin{equation}
\label{param}
    f_0=\frac{Fl_0^3}{C_{0}\Gamma_{a}D_a}\,,
        \quad
    \alpha_0=\frac{D_b}{D_a}\,,
        \quad
    \delta_0=\frac{\Gamma_a-\Gamma_b}{\Gamma_a} \,,
\end{equation}
which respectively relate to the flux, diffusion anisotropy and step
stiffness difference.

\begin{figure}
\centering
\includegraphics[width=0.5\textwidth]{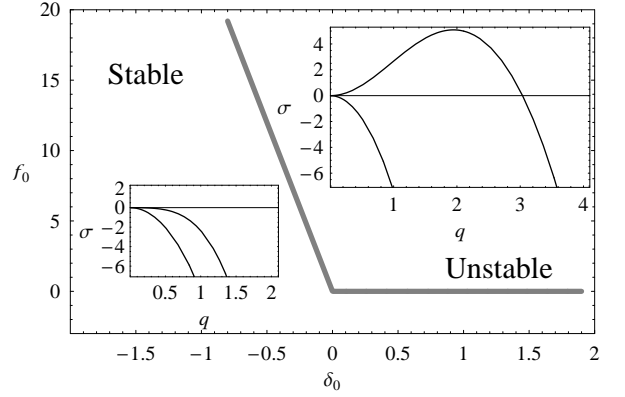}
\caption{Stability diagram in the plane $(f_0,\delta_0)$, for
$\alpha_0>1$. The thick gray line $f_0=f_{0c}$ given by
Eq.~(\protect\ref{fc}), separates the unstable region (right side),
from the stable one (left side). The dispersion relation $\sigma(q)$
with its two branches is shown in each region.} \label{fig2}
\end{figure}

We investigate now the linear stability of a train of equidistant
steps traveling at a constant velocity $f_0$ when perturbed
transversally. The shape of the steps can be decomposed in Fourier
modes of the form $x_a^n(y,t)=f_0 t+2n+\xi_{a}^n(y,t)$ and
$x_b^n(y,t)=f_0 t+2n+1+\xi_{b}^n(y,t)$ with $\xi_a^n$ and $\xi_b^n$
the perturbation amplitudes varying as $\exp(\sigma(q,\phi) t+i q
y+i n \phi)$, where $q$ is the wavenumber and $\phi$ the phase (see
Fig.~\ref{fig1}). Inserting these expressions into
(\ref{bcf1})-(\ref{vab2}), we obtain the general dispersion relation
$\sigma=\sigma(q,\phi)$. The dispersion relation possesses, for each
$\phi$, two branches corresponding to a stable $\sigma_s$ and an
unstable $\sigma_u$ mode. The maximum growth rate is reached for the
in-phase perturbation $\phi=0$ (Fig.~\ref{fig2}). In the following
we consider only this in-phase mode, thus neglecting the $n$
dependence $\xi_a^n=\xi_a$, and $\xi_b^n=\xi_b$ (the system is
periodic in the $x$-direction with period $2l_0$). We find that the
instability appears above a flux threshold $f_{0c}$ given by,
\begin{equation}
    \label{fc}
    f_0>f_{0c}=-\frac{12 \alpha_0 \delta_0}{\alpha_0-1} \, ,
\end{equation}
where $f_0>0$. The stability domain in the plan $(f_0,\delta_0)$ is
shown in Fig.~\ref{fig2}. Typically the instability is related to a
large diffusion $D_b$ on terrace $b$ ($\alpha_0\gg1$) accompanied by
a small stiffness $\tilde\gamma_b$ of step $S_b$ ($\delta_0>0$).

Although the full expression of the dispersion relation is
cumbersome, near the instability threshold we can introduce a small
parameter $\epsilon$ measuring the distance to the threshold. This
parameter arises naturally when considering the long wavelength
limit, in which $q\rightarrow\epsilon q$. In this limit the relevant
scaled parameters are chosen to be:
\begin{equation}
\label{scale}
    f_0=\epsilon f,\;
    \delta_0=\epsilon \delta,\;
    \alpha_0=1+\epsilon^2 \alpha\,,
\end{equation}
together with a rescaling of the stiffness
$\Gamma_a\rightarrow\epsilon^2\Gamma_a$. This scaling will lead to a
consistent weak nonlinear expansion, as it will be shown below. To
lowest order the two branches of $\sigma=\sigma(q,\phi)$ are,
\begin{eqnarray}
\label{ss}
    \sigma_s &=& -4\epsilon^4 q^2 \,, \\
\label{su}
    \sigma_u &=& \frac{\epsilon^6}{48}
        f\alpha (f\alpha+12\delta)q^2-\epsilon^6 q^4\,.
\end{eqnarray}
The growth rate $\sigma_u$ is maximum for the wavenumber
$q_{m}=[f\alpha(f\alpha+12\delta)]^{1/2}/4\sqrt{6}$. Even for
vanishing $\delta$ (no difference between step stiffness) an
instability is still present driven by diffusion anisotropy and
growth effect, $\sigma_u(q_m)\sim f^4\alpha^4$.

Taking for example typical values for Si(001) at two different
temperatures $T=1000\,\mathrm{K}$ and $T=800\,\mathrm{K}$ (with
$l_0=2\,10^{-8}\,\mathrm{m}$ and $F=0.1\,\mathrm{ML\,s^{-1}}$),
$f_0=0.3$, and $\,4.3$, $\alpha_0=45$, and $\,119$, and
$\delta_0=0.9$, we obtain that the typical wavelength of the
meanders is of the order of $250\,\mathrm{nm}$ and
$60\,\mathrm{nm}$, respectively. These sizes are in the range of the
transverse modulations (ripples) of the step bunches observed in the
experiments {\cite{schelling99,schelling00,pascale03,myslivecek02b}.

We study at present the nonlinear evolution of the meandering
instability, in the limit of weak amplitudes and long wavelengths.
An inspection of Eqs.~(\ref{ss})-(\ref{su}) for the damping and
growth rates, suggests that we should consider different time scales
and amplitudes for the stable and unstable branches. We introduce
the functions $s(y,t)$ and $u(y,t)$ corresponding to the amplitudes
of the stable and the unstable branches respectively. These
amplitudes are related to the step shape by,
\begin{equation}
    \left(\begin{array}{c}
        \xi_a\\
        \xi_b\end{array}\right)=\epsilon
        \mathcal{M}_0(\epsilon)
    \left(\begin{array}{c}\epsilon\, s\\
        u\end{array}\right)\,,
\end{equation}
where $\mathcal{M}_0$ is the matrix formed with the eigenvectors
associated to the linear dispersion relation and depends on the
physical parameters $(f_0,\alpha_0,\delta_0)$. In order to obtain
the relevant nonlinear dynamics we use a standard multiscale method.
Adatoms concentrations and step shapes are expanded in powers of
$\epsilon$. The amplitudes of this expansion depend on the slowly
varying space $\epsilon y$ and two time variables $\epsilon^4 t$ and
$\epsilon^6 t$. In particular the stable and unstable shape
functions are given by $s=s(\epsilon^4 t,\epsilon y)$ and
$u=u(\epsilon^6 t,\epsilon y)$. Solving diffusion equations
(\ref{bcf1}-\ref{bcf2}) and boundary conditions
(\ref{bc1}-\ref{bc2}) up to order $\epsilon^7$, and inserting the
results into the step velocity equations (\ref{vab1}-\ref{vab2}) we
find the equation for the unstable mode:
\begin{equation}\label{CKS}
    \partial_t u=-\partial^2_y\left[
        \frac{f\alpha }{48}(f\alpha+12\delta) u+
        \partial^2_y u+
        \frac{f}{12}(\partial_y u)^2
    \right]\,,
\end{equation}
where we renamed the slow variables $\epsilon^6 t\rightarrow t$ and
$\epsilon y\rightarrow y$. After rescaling we can write
Eq.~(\ref{CKS}) in the form $\partial_t u = -\partial^2_y[u +
\partial^2_{y}u + (\partial_y u)^2/2]$. This is a conserved version of the
Kuramoto-Sivashinsky equation (CKS) which also describes the weak
nonlinear regime of the step bunching instability \cite{gillet01}.
Numerical simulations (performed with an extra term $\partial^3_y
u$) have revealed a non interrupted coarsening dynamics, the
characteristic size of coalesced step bunches increases as $t^{1/2}$
\cite{gillet01}. In our context the $\partial^3_y u$ term,
introducing a dispersive drift, is absent, but this should not
change the $t^{1/2}$ scaling. Our direct simulations confirmed this
behavior and also demonstrate a linear time growth of the
characteristic meander amplitude $\langle u^2\rangle^{1/2}\sim t$
(spatial average is denoted $\langle\cdots\rangle$). A typical
spatiotemporal evolution from a random initial condition is shown in
Fig.~\ref{fig3}. In the context of Bales-Zangwill meandering
instability, it was shown that the dynamics of steps is fully
nonlinear, excluding CKS equation, although it would be compatible
with the basic symmetries of the system \cite{gillet00}.

Simple similarity and matching arguments lead to a complete picture
for the long time behavior of (\ref{CKS}). It is worth noting that
the CKS equation admits an exact particular solution in the form of
a stationary parabola $u(y,t)=-y^2/2$. We also note that, for
rapidly decreasing or bounded functions, the moment of order one of
$u(y,t)$ is conserved while the second order moment satisfies
\begin{equation}\label{mom}
    \frac{d}{dt}\frac{1}{2}\int u^2 dy=
        \int [(\partial_y u)^2-(\partial^2_y u)^2] dy\,,
\end{equation}
showing that the amplitude of $u$ tends to increase in regions where
the gradient term $\partial_y u$ inside the integral dominates the
curvature term $\partial^2_y u$. This suggests that the dynamics of
large amplitude smooth regions of $u(y,t)$ is almost independent of
the four derivative (stabilizing) term in (\ref{CKS}). Trying a
similarity solution
\begin{equation}\label{sim}
    u=t^a\varphi(y/t^b)\,,\quad Y=y/t^b
\end{equation}
of $\partial_t u = -\partial^2_y[u + (\partial_y u)^2/2]$, we find
immediately the exponents $b=1/2$ and $a=1$, which agree with our
numerical results. The fourth order derivative term behaves as
$\partial^4_y u\sim \partial^4_Y\varphi/t$, which is consistent with
the above assumption that it should be negligible in the considered
regime. Moreover, the similarity equation for $\varphi$:
$\varphi-(Y/2)\partial_Y\varphi +\partial^2_Y[\varphi+
(\partial_Y\varphi)^2/2]=0$ has a solution in the form of a bounded
parabola:
\begin{equation}\label{parabola}
    u(y,t)=-y^2/2,\quad |y|<y_0(t)
\end{equation}
and zero elsewhere. The parabola edge $y_0(t)$ is determined by the
condition
\begin{equation}
    \frac{d}{dt}\frac{1}{4}\int_0^{y_0(t)} y^4 dy=
        2\int_0^{y_0(t)}y^2 dy\,,
\end{equation}
a consequence of (\ref{mom}), which gives $y_0(t)=(16t/3)^{1/2}$.
The general, asymptotic solution of (\ref{CKS}) can be thought as a
superposition of parabolas satisfying (\ref{parabola}) (see
Fig.~\ref{fig3}). The joining region between the parabolas possesses
a high curvature and can be described by the reduced inner equation
$\partial^2_yu(y)+(1/2)(\partial_yu(y))^2=2k^2=\mathrm{const.}$,
whose solution is of the form $u(y)=2\log[\cosh k(y-y_0(t))]$.
Matching with the outer solution (\ref{parabola}) we find that
$k=y_0(t)/2\sim\sqrt{t}$. Therefore, the curvature of the joining
line increases as $\kappa\sim t$.

\begin{figure}
\centering
\includegraphics[width=0.48\textwidth]{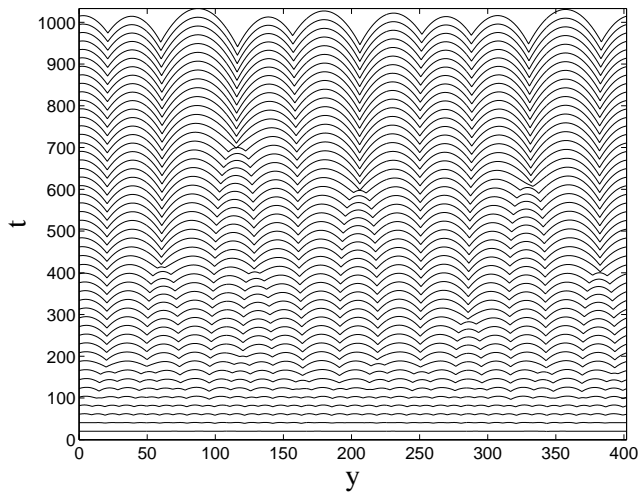}
\caption{Spacetime plot of $u(y,t)$ with nondimensional $y$ and $t$,
given by the numerical solution of the CKS equation. The
coarse-graining of structures leads to a superposition of parabolas,
with a size $\langle u^2\rangle^{1/2}\sim t$. In the long time state
all the parabolas tend to have unity curvature at their maximum, and
width increasing as $\sqrt{t}$.} \label{fig3}
\end{figure}

In this Letter, we have shown that the effect of step stiffness
difference and diffusion anisotropy induces a meandering instability
during surface growth. We  have first presented a linear stability
analysis of a train of steps using a simplified two-dimensional
extension of the model studied in Ref.~\cite{frisch05}. We have
shown that the interplay between diffusion anisotropy and the step
stiffness effect under growth conditions leads to a finite
wavelength instability which is maximum for the in-phase mode. Our
results are complemented by a weak nonlinear analysis of the step
dynamics which reveals that the amplitude of the meanders is
governed by the conserved Kuramoto-Sivashinsky equation (CKS) which
displays non-interrupted coarsening.  We believe that the morphology
observed in experiments of molecular beam epitaxy on Si(001)
slightly disoriented towards the $[110]$ direction, reported in
Refs.~\cite{schelling99,schelling00,pascale03,myslivecek02b}, can be
explained by the nonlinear evolution of the step bunching and step
meandering instabilities arising simultaneously. Indeed, we will
present elsewhere an investigation of the two-dimensional dynamics
originated by the nonlinear coupling between these kinetic effects,
and we will discuss their role in the formation of the ripples shown
for instance in Fig.~1 of Ref.~\cite{schelling00}.

\acknowledgments{We would like to thank Jean-Noël Aqua, Isabelle
Berbézier, Matthieu Dufay, and Chaouqi Misbah for stimulating
discussions.}

\bibliography{mondocument1}

\end{document}